\documentclass[twocolumn,prl,showpacs,
floatfix,preprintnumbers,nofootinbib]{revtex4}

\usepackage{epsfig}
\usepackage{amsmath,bm,color}

\newcommand{\ph}{\phantom{{-}1}}

\begin{document}

\title{Axial symmetry breaking in self-induced flavor conversion of
supernova neutrino fluxes}

\author{Georg Raffelt, Srdjan Sarikas and David de Sousa Seixas}
\affiliation{Max-Planck-Institut f\"ur Physik
(Werner-Heisenberg-Institut), F\"ohringer Ring 6, 80805 M\"unchen,
Germany}

\date{31 May 2013, revised 17 July 2013, Typographical errors (in red) corrected 17 Nov.~2014}


\begin{abstract}
Neutrino-neutrino refraction causes self-induced flavor conversion
in dense neutrino fluxes. For the first time, we include the azimuth
angle of neutrino propagation as an explicit variable and find a new
generic multi-azimuth-angle (MAA) instability which, for simple
spectra, occurs in the normal neutrino mass hierarchy. Matter
suppression of this instability in supernovae requires larger
densities than the traditional bimodal case. The new instability
shows explicitly that solutions of the equations for collective
flavor oscillations need not inherit the symmetries of initial or
boundary conditions. This change of paradigm requires
reconsideration of numerous results in this field.
\end{abstract}

\preprint{MPP-2013-127}


\maketitle


{\em Introduction}.---Flavor oscillations depend strongly on matter
because the weak-interaction potential $\sqrt{2}\,G_{\rm F}n_e$ can
far exceed the oscillation energy $\omega=\Delta
m^2/2E$~\cite{Wolfenstein:1977ue, Kuo:1989qe}. A matter gradient can
cause complete flavor conversion (MSW effect), notably for neutrinos
streaming from a supernova (SN) core \cite{Mikheev:1986gs,
Dighe:1999bi, Serpico:2011ir}.  In addition, the large neutrino flux
itself causes strong neutrino-neutrino refraction
\cite{Pantaleone:1992eq} and can lead to self-induced flavor
conversion \cite{Sawyer:2005jk, Duan:2005cp, Duan:2006an,
Duan:2010bg}. This effect is very different from MSW conversion
because the flavor content of the overall ensemble remains fixed.
Instead, flavor is exchanged between different momentum modes and
can lead to interesting spectral features~\cite{Duan:2006an,
Duan:2010bg, Raffelt:2007cb, Duan:2007fw, Fogli:2007bk,
Dasgupta:2009mg, Cherry:2013mv}. Self-induced flavor conversion can
become large because of self-amplification within the interacting
neutrino system, which in turn requires instabilities (collective
run-away solutions) in flavor space~\cite{Sawyer:2008zs,
Banerjee:2011fj, Sarikas:2011am, Sarikas:2012ad}.

Our main point is that run-away solutions need not inherit the
symmetries of initial or boundary conditions. For self-induced
flavor conversion in SNe, global spherical symmetry was always
assumed and therefore axial symmetry in every direction. However,
our linearized stability analysis shows that local axial symmetry
can be broken by a multi-azimuth angle (MAA) instability. For simple
spectra it arises in the normal hierarchy (NH) of neutrino masses,
whereas the traditional bimodal instability \cite{Samuel:1995ri,
Duan:2005cp, Hannestad:2006nj} occurs in the inverted hierarchy
(IH).

Core-collapse SNe show large convective overturns, the standing
accretion shock instability, or simply rotation. However, our new
effect is not caused by the concomitant asymmetries of neutrino
emission, but by the intrinsic flavor instability of an axially
symmetric neutrino flux, an effect which does not strongly depend on
the exact azimuth distribution of emission.

In the early universe, one can integrate out the factor $1-{\bf
v}{\cdot}{\bf v}'$ from the current-current neutrino interaction and
then finds the bimodal instability \cite{Samuel:1995ri}. However,
for equal neutrino and anti-neutrino densities it was found that
allowing angle modes to evolve independently enables run-away
solutions in both hierarchies \cite{Raffelt:2007yz}. In this early
study it was not recognized that such multi-angle instabilities are
far more general.

Two-flavor neutrino-neutrino refraction can be written in the form
of the spin-pairing Hamiltonian that appears in many areas of
physics~\cite{Pehlivan:2011hp}. When all flavor spins interact with
each other with the same strength, this Hamiltonian has as many
invariants as variables and thus is integrable, explaining the
N-mode coherent solutions \cite{Pehlivan:2011hp, Yuzbashyan:2008,
Raffelt:2010za, Raffelt:2011yb}. After including the factor $1-{\bf
v}{\cdot}{\bf v}'$, these simple properties are probably lost. It
would be interesting to study this multi-angle spin-pairing
Hamiltonian to develop a deeper mathematical understanding of our
system.

Our more modest goal here is to prove explicitly the existence of
the MAA instability in the simplest SN setting and how it is
affected by matter, allowing for a first understanding of the MAA
effect.

{\em Equations of motion}.---We describe neutrinos by $3{\times}3$
flavor matrices $\varrho(t,{\bf r},E,{\bf v})$, where the diagonal
elements are occupation numbers for $\nu_e$, $\nu_\mu$ and
$\nu_\tau$ while the off diagonal elements encode correlations
caused by flavor oscillations. We use negative $E$ to denote
$\bar\nu$ in which case $\varrho$ includes a minus sign: the
diagonal elements are negative $\bar\nu$ occupation numbers. (One
needs \hbox{$6{\times}6$} matrices to include $\nu$--$\bar\nu$
correlations that could arise from novel lepton-number violating
interactions \cite{Sawyer:2010jk, Volpe:2013uxl} or from Majorana
spin-flavor oscillations \cite{Dvornikov:2011dv,deGouvea:2012hg}.)

In the absence of collisions, neutrino propagation is described by
the Liouville equation~\cite{Sigl:1992fn, Cardall:2007zw}
\begin{equation}
\left(\partial_t
+ {\bf v}{\cdot} \nabla_{\bf r}
\right)
\varrho
= - i \, \left[\,
{\sf H}\,,\, \varrho\,
\right]\,,
\end{equation}
where $\varrho$ and ${\sf H}$ are functions of $t$, ${\bf r}$, $E$
and ${\bf v}$. (Except for $\varrho$, we use capital sans-serif
letters to denote matrices in flavor space.) The Hamiltonian matrix
is
\begin{equation}\nonumber
{\sf H}=
\frac{{\sf M}^2}{2E}
+ \sqrt{2}\,G_{\rm F}\!\left[{\sf N}_\ell
+ \int_{-\infty}^{+\infty}\hskip -1em dE'\, {\color{red}E'^2}\!\!
\int\frac{d{\bf v}'}{(2\pi)^3}\,
  \varrho'(1{-}{\bf v}{\cdot}{\bf v}')\right],
\end{equation}
where ${\sf M}^2$ is the matrix of mass-squares, causing vacuum
oscillations. The matrix of charged-lepton densities, ${\sf N}_\ell$,
includes the background matter effect. The $d{\bf v}'$ integral is
over the unit sphere and $\varrho'$ depends on $t$, ${\bf r}$, $E'$
and ${\bf v}'$.

In general, this is an untractable $7$-dimensional problem. As a
simplification we assume stationarity and drop the time dependence.
We also assume spherically symmetric emission, but no longer enforce
local axial symmetry of the solution. We still assume that
variations in the transverse direction are small so that $\varrho$
depends only on $r$, $E$ and ${\bf v}$. In other words, we study
neutrino propagation only in the neighborhood of a chosen location
and do not worry about the global solution.

We consider neutrinos that stream freely after emission at some
fiducial inner boundary $R$ (``neutrino sphere''). If we describe
neutrinos by their local ${\bf v}$, the zenith range of occupied
modes depends on radius. To avoid this effect we use instead the
emission angle $\vartheta_R$ to label the modes. The variable
$u=\sin^2\vartheta_R$ is even more convenient because blackbody-like
isotropic emission at $R$ corresponds to a uniform distribution on
$0\leq u\leq1$. The radial velocity of a mode $u$ at radius $r$ is
$v_{r,u}=(1-u\,R^2/r^2)^{1/2}$ and the transverse velocity is
$\beta_{r,u}=u^{1/2}\,R/r$.

To study quantities that evolve only as a consequence of flavor
oscillations, we introduce flux matrices~\cite{EstebanPretel:2008ni}
by
\begin{equation}
\frac{{\sf F}(r,E,u,\varphi)}{4 \pi r^2}\,\frac{dE\,du\,d\varphi}{v(u,r)}=
{\varrho}(r,{\bf p}) \frac{d^3{\bf p}}{(2\pi)^3}\,,
\end{equation}
where $\varphi$ is the azimuth angle of ${\bf v}$. The Liouville
equation finally becomes $\partial_r {\sf F}=-i\,[\,{\sf H},{\sf
F}\,]$, the vacuum and matter terms receive a factor $v^{-1}$, and
the $\nu$--$\nu$ part is
\begin{equation}
{\sf H}_{\nu\nu}=
\frac{\sqrt{2}\,G_{\rm F}}{\color{red}4\pi r^2}\int d\Gamma'\,
{\sf F}'\,\,\frac{1-vv'-{\bm\beta}{\cdot}{\bm\beta}'}{vv'}\,,
\end{equation}
where $\int
d\Gamma'=\int_{-\infty}^{+\infty}dE'\int_0^1du'\int_0^{2\pi}d\varphi'$.
In addition, we find ${\bm\beta}{\cdot}{\bm\beta}'= \sqrt{u
u'}\,(R^2/r^2)\,\cos(\varphi-\varphi')$. Enforcing axial symmetry
would remove the ${\bm\beta}{\cdot}{\bm\beta}'$ term, and this is what
was done in the previous literature.

{\em Two flavors.}---Henceforth we consider only two flavors $e$ and
$x=\mu$ or $\tau$ and describe energy modes by $\omega=\Delta
m^2/2E$. We write the $2{\times}2$ flux matrices in the form
\begin{equation}
{\sf F}=\frac{{\rm Tr}\,{\sf F}}{2}
+\frac{F_e^R-F_x^R}{2}\,
\begin{pmatrix}
s&S\\
S^*&-s
\end{pmatrix}\,,
\end{equation}
where $F_{e,x}^R(\omega,u,\varphi)$ are the flavor fluxes at the
inner boundary radius $R$. All other quantities depend on $r$,
$\omega$, $u$ and $\varphi$. The flux summed over all flavors, ${\rm
Tr}\,{\sf F}$, is conserved and can be ignored in commutators. The
$\nu_e$ survival probability, $\frac{1}{2}(1+s)$, is given in terms
of what we call the swap factor $-1\leq s\leq1$. The off diagonal
element $S$ is complex and $s^2+|S|^2=1$.

We introduce the dimensionless spectrum $g(\omega,u,\varphi)$,
representing $F_e^R-F_x^R$. It is negative for antineutrinos where
$\omega<0$, and normalized to the $\bar\nu$ flux, i.e.,
$\int_{-\infty}^0 d\omega\int_0^1 du
\int_0^{2\pi}d\varphi\,g(\omega,u,\varphi)=-1$. The $\nu$--$\bar\nu$
asymmetry is $\epsilon=\int d\Gamma\,g$ where $\int d\Gamma=
\int_{-\infty}^{+\infty}d\omega\int_0^1du\int_0^{2\pi}d\varphi$.

Refractive effects are provided by the $r$-dependent parameters
\cite{Banerjee:2011fj}
\begin{eqnarray}
\lambda&=&\sqrt{2}\,G_{\rm F}\,\left[n_e(r)-n_{\bar e}(r)\right]\,\frac{R^2}{2r^2}\,,
\nonumber\\
\mu&=&\frac{\sqrt{2}\,G_{\rm F}\,[F_{\bar\nu_e}(R)-F_{\bar\nu_x}(R)]}{4\pi r^2}
\,\frac{R^2}{2r^2}\,.
\end{eqnarray}
In analogy to $g$, we normalize the effective $\nu$--$\nu$
interaction energy $\mu$ to the $\bar\nu_e$--$\bar\nu_x$ flux
difference at $R$. The factor $R^2/2r^2$ highlights that only the
multi-angle impact of refraction is relevant \cite{Banerjee:2011fj}.

So finally the stability analysis uses the spectrum
$g(\omega,u,\varphi)$, the effective $\nu$--$\nu$ interaction energy
$\mu\propto r^{-4}$, and the total matter effect parameterized by
$\bar{\lambda}=\lambda +\epsilon\mu$. For $\Delta m^2>0$, our
equations correspond to IH, whereas NH can be implemented with
\hbox{$\Delta m^2\to-\Delta m^2$} or equivalently via
$\omega\to-\omega$ in the vacuum term of ${\sf H}$.

{\em Linearized stability analysis.}---At high density, neutrinos
are produced in flavor eigenstates and propagate as such until the
initially small off diagonal elements of ${\sf F}$ grow large. This
can happen by an MSW resonance, which in SNe typically occurs at
much larger distances than self-induced conversions. In the latter
case, which we study here, the sudden growth is caused by an
exponential run-away solution. We assume that no such instability
occurs out to $r\gg R$, so we use the large-distance approximation
where the transverse neutrino velocity is small. To linear order in
$S$, we have $s=1$ and find
\begin{eqnarray}\label{eq:smallEoM}
i\partial_r S&=&
(\omega + u\bar\lambda)\,S\\
&-&\mu \int d\Gamma'\, \left[u+u'-2\sqrt{uu'}\cos(\varphi-\varphi')\right]\,g'\,S'\,.
\nonumber
\end{eqnarray}
We write solutions as
$S(r,\omega,u,\varphi)=Q_\Omega(\omega,u,\varphi)\,e^{-i\Omega r}$
with complex eigenfrequency $\Omega=\gamma+{\rm i}\kappa$ and
eigenvector $Q_\Omega(\omega,u,\varphi)$, which satisfy the eigenvalue equation
\begin{eqnarray}\label{fourier-eom}
(\omega + u\bar\lambda - \Omega)\,Q_\Omega&=&
\nonumber\\
&&\kern-9em
\mu \int d\Gamma'\,\left[u+u'-2\sqrt{uu'}\cos(\varphi-\varphi')\right]\,g'\,Q_\Omega'\,.
\end{eqnarray}
The rhs has the form $a + bu+\sqrt{u}\,(c \cos\varphi+d\sin\varphi)$
with complex numbers $a$, $b$, $c$ and $d$, so the eigenvectors are
\begin{equation}
Q_\Omega=\frac{a + bu +\sqrt{u}(c\cos\varphi+d\sin\varphi)}{\omega+u\bar\lambda-\Omega}\,.
\label{eq:Eigenfunctions}
\end{equation}
After inserting Eq.~(\ref{eq:Eigenfunctions}) into
(\ref{fourier-eom}), self-consistency requires
\begin{equation}\label{eq:abmatrix}
\begin{pmatrix}
   I_1{-}1         &    I_2             &    I_{3/2}^{\rm c}      &      I_{3/2}^{\rm s} \\
   I_0             &    I_1{-}1         &    I_{1/2}^{\rm c}      &      I_{1/2}^{\rm s} \\
-2 I_{1/2}^{\rm c} & -2 I_{3/2}^{\rm c} & -2 I_{1}  ^{\rm cc}{-}1 &   -2 I_{1}  ^{\rm sc}\ph \\
-2 I_{1/2}^{\rm s} & -2 I_{3/2}^{\rm s} & -2 I_{1}  ^{\rm sc}\ph  &   -2 I_{1}  ^{\rm ss}{-}1
\end{pmatrix}
\begin{pmatrix} a \\ b \\ c \\ d \end{pmatrix}=0\,,
\end{equation}
where
\begin{equation}
I_n^{\rm c(s)}=\mu \int du\,d\omega\,d\varphi\,
\frac{u^n\,g(\omega,u,\varphi)}{\omega+u\bar\lambda-\Omega}\cos\varphi\,(\sin\varphi)\,.
\label{eq:In-def}
\end{equation}
Nontrivial solutions exist if the determinant of the matrix
vanishes. The mass hierarchy IH $\to$ NH is changed by
$\omega\to-\omega$ in the denominator of Eq.~(\ref{eq:In-def}).

{\em Axial symmetry of neutrino emission.}---As a next step, we
simplify to $g(\omega,u,\varphi)\to g(\omega,u)/2\pi$.  Now only the
$\varphi$ integrals with $\sin^2\varphi$ and $\cos^2\varphi$ survive
and yield $I^{\rm cc}_1=I^{\rm ss}_1=\frac{1}{2}I_1$, leaving us
with
\begin{equation}
\label{eq:block_matrix}
\begin{pmatrix}
I_1{-}1 & I_2     & 0                    & 0  \\
I_0     & I_1{-}1 & 0                    & 0  \\
0       & 0       & -(I_1{+}1) & 0  \\
0       & 0       & 0          & -(I_1{+}1)
\end{pmatrix}
\begin{pmatrix} a \\ b \\ c \\ d \end{pmatrix}=0\,.
\end{equation}
This system has nontrivial solutions if
\begin{equation}\label{eq:determinant}
(I_1-1)^2=I_0I_2 \quad\hbox{or}\quad I_1 = -1\,,
\end{equation}
where the integral expressions are the same as in the previous
azimuthally symmetric case \cite{Banerjee:2011fj}.

The first equation corresponds to nontrivial solutions for $a$ and
$b$ and yields the instabilities found in previous works. In IH this
is the well-known bimodal solution, present even for the
single-angle case of only one zenith mode. In NH the bimodal
solution does not exist and multi-angle effects are necessary for
any run-away solution. For a nontrivial distribution of zenith
angles, the first equation leads to a solution
\cite{Banerjee:2011fj} which we now denote the multi-zenith angle
(MZA) instability.

The second equation allows for nonzero $c$ and $d$, providing
solutions with nontrivial $\varphi$ dependence, unstable only in NH.
The previous solutions remain unaffected by MAA, whereas in NH new
solutions appear.

These cases become more explicit if we ignore matter
($\bar\lambda=0$) and assume the spectrum factorizes as
$g(\omega,u)\to g(\omega)h(u)$. With $I=\mu\int
d\omega\,g(\omega)/(\omega-\Omega)$, Eq.~(\ref{eq:determinant}) is
\begin{equation}\label{eq:determinant2}
I^{-1}=q_{j}=\begin{cases}
\langle u\rangle+\langle u^2\rangle^{1/2}&\hbox{for~~$j={}$bimodal}\,,\\
\langle u\rangle-\langle u^2\rangle^{1/2}&\hbox{for~~$j={}$MZA}\,,\\
-\langle u\rangle&\hbox{for~~$j={}$MAA}\,\\ \end{cases}
\end{equation}
Note that $q_j$ is positive in the first case, and negative in the
second and third. In IH, the first case is the only one providing an
instability (bimodal) and exists for any $u$ distribution. In NH,
the first case is always stable, while the second case yields the
MZA solution. It does not exist for single angle where $\langle
u\rangle=\langle u^2\rangle^{1/2}$. The third case  exists for any $u$ distribution. For simple (single-crossed) spectra, it
provides the new MAA solution only in NH.

We illustrate these findings with a simple example and consider
single neutrino energy ($\omega=\pm\omega_0$), i.e., the spectrum
$g(\omega)=-\delta(\omega{+}\omega_0)+(1{+}\epsilon)\,\delta(\omega{-}\omega_0)$.
Equation~(\ref{eq:determinant2}) is now equivalent to the quadratic
equation
\begin{equation}
\frac{\omega_0^2-\Omega^2}{2\omega_0+\epsilon(\omega_0+\Omega)}=\mu\,q_j\equiv\mu_j\,,
\end{equation}
where $j={}$bimodal, MZA or MAA. The solutions are
\begin{equation}
\Omega_j= \frac{1}{2}\left(-\epsilon\mu_j\pm
\sqrt{ (2\omega_0-\epsilon\mu_j)^2 - 8 \omega_0\mu_j} \right)\,.
\end{equation}
Exponentially growing solutions ($\kappa={\rm Im}\,\Omega>0$) can
only exist when $\omega_0\mu_j>0$. IH corresponds to $\omega_0>0$
and only the first case has $q_j>0$, providing the bimodal
instability. NH corresponds to $\omega_0<0$ so that the second and
third cases provide the MZA and MAA instabilities.

The system is unstable for $\mu_j$ between the limits
$2\omega_0/(\sqrt{1+\epsilon}\pm1)^2$. The maximum growth rate
obtains for $\mu_j=2\omega_0 (2+\epsilon)/\epsilon^2$ and is
$\kappa_{\rm max}=2|\omega_0|\,\sqrt{1+\epsilon}/\epsilon$.
Therefore, a typical growth rate is a few times the vacuum
oscillation frequency. For $\epsilon=1/2$ we find $\kappa_{\rm
max}=2\sqrt{6}\,|\omega_0|\approx4.90\,|\omega_0|$.

{\em Azimuth distribution.}---According to the expression for the
eigenfunction $Q_\Omega$ in Eq.~(\ref{eq:Eigenfunctions}), the off
diagonal elements of the $\varrho$ matrices develop an exponentially
growing ``dipole term'' $c\,\cos(\varphi)+d\,\sin(\varphi)$, which
represents an ellipse in the complex plane. Its orientation and
ellipticity is chosen by some initial disturbance. If neutrino
emission is not axially symmetric, it provides a macroscopic seed,
but otherwise the situation is largely the same.

In this sense, our main point is that the linearized system supports
run-away solutions where the exponentially growing off diagonal
$\varrho$ elements depend on $\varphi$ even if the diagonal
elements, represented by $g(\omega,u,\varphi)$, do not depend on
$\varphi$ because of axially symmetric emission.

If we represent the $\varphi$ dependence by $N$ discrete angles
$\varphi_i$ with $i=1,\dots,N$, the corresponding distributions
$\delta(\varphi-\varphi_i)$ can be expanded in terms of functions
$\cos(n\varphi)$ and $\sin(n\varphi)$. One can then show that the
linearity of the eigenfunctions $Q_\Omega$ in $\cos(\varphi)$ and
$\sin(\varphi)$ implies that no new instabilities arise in the
discretized system. No spurious instabilities appear, in contrast to
discrete zenith angles \cite{Sarikas:2012ad}, where the
eigenfunctions depend on $u$ in nonlinear ways.

\begin{figure}
\includegraphics[width=0.85\columnwidth]{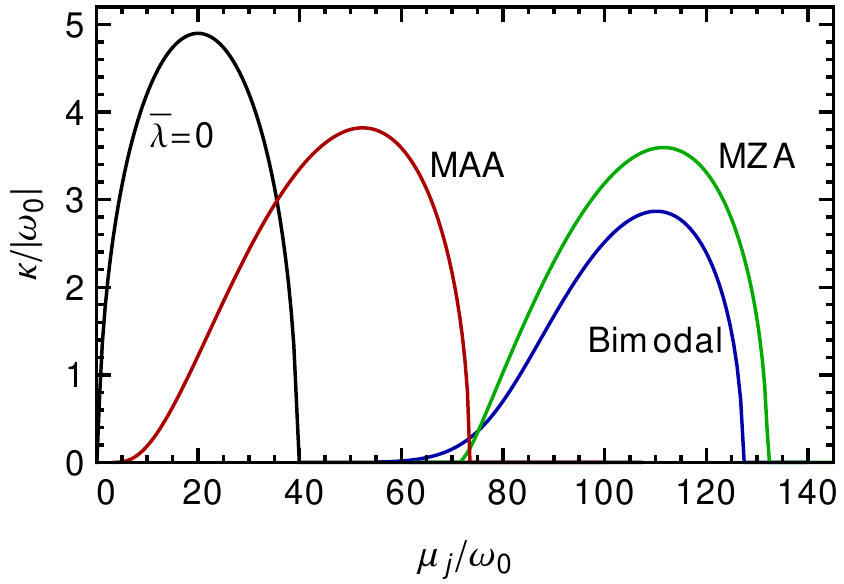}
\caption{Growth rate $\kappa$ for blackbody-like zenith
distribution, single energy $\pm\omega_0$, and $\epsilon=1/2$.
{\em Black line:} All cases for $\bar\lambda=\lambda+\epsilon\mu=0$
(no matter effect).
{\em Other lines:} Indicated unstable cases for
$\bar\lambda=300\,|\omega_0|$.\label{fig:matter}}
\end{figure}

{\em Impact of matter.}---If there is only one zenith angle, matter
has no impact on $\kappa$ because $\bar\lambda u$ in the resonance
denominator simply shifts the real part of $\Omega$. In general,
$\bar\lambda u$ is different for every zenith angle trajectory, along
which neutrinos acquire different matter-induced phases. If
$\bar\lambda$ is large, the unstable region shifts to larger
$\mu$-values \cite{Banerjee:2011fj} as shown in Fig.~\ref{fig:matter}
for all three cases. We have used blackbody-like zenith distribution
(uniform on \hbox{$0\leq u\leq1$}) where $q_{\rm
bimodal}=1/2+1/\sqrt3\approx1.077$, $q_{\rm
MZA}=1/2-1/\sqrt3\approx-0.077$ and $q_{\rm MAA}=-1/2$. On the
horizontal axis in Fig.~\ref{fig:matter}, we use $\mu_j$ as a
variable, so the physical $\mu$ range is very different for the three
cases.

Numerically it appears that for large $\bar\lambda$, the instability
occurs for $\alpha_j\mu \sim \bar\lambda$, where $\alpha_j$ is a
coefficient different for each case. It also appears that for the
bimodal and MZA cases, actually $\alpha_j\sim|q_j|$ and we roughly
have $\bar\lambda\sim|\mu_j|$. Note that
$\bar\lambda=\epsilon\mu+\lambda\sim |q_{\rm MZA}|\mu=0.077\,\mu$, so
that, for reasonable values of $\epsilon$, the matter density
$\lambda$ would have to be negative---the MZA instability is
self-suppressed by the unavoidable effect of neutrinos themselves,
and plays no role in a realistic SN situation. On the other hand, we
find the new MAA instability the least sensitive to matter effects,
as the instability region shifts only for much larger interaction
strength ($\alpha_{\rm MAA} \sim 6\, |q_{\rm MAA}|$).

\begin{figure}[t]
\includegraphics[width=0.93\columnwidth]{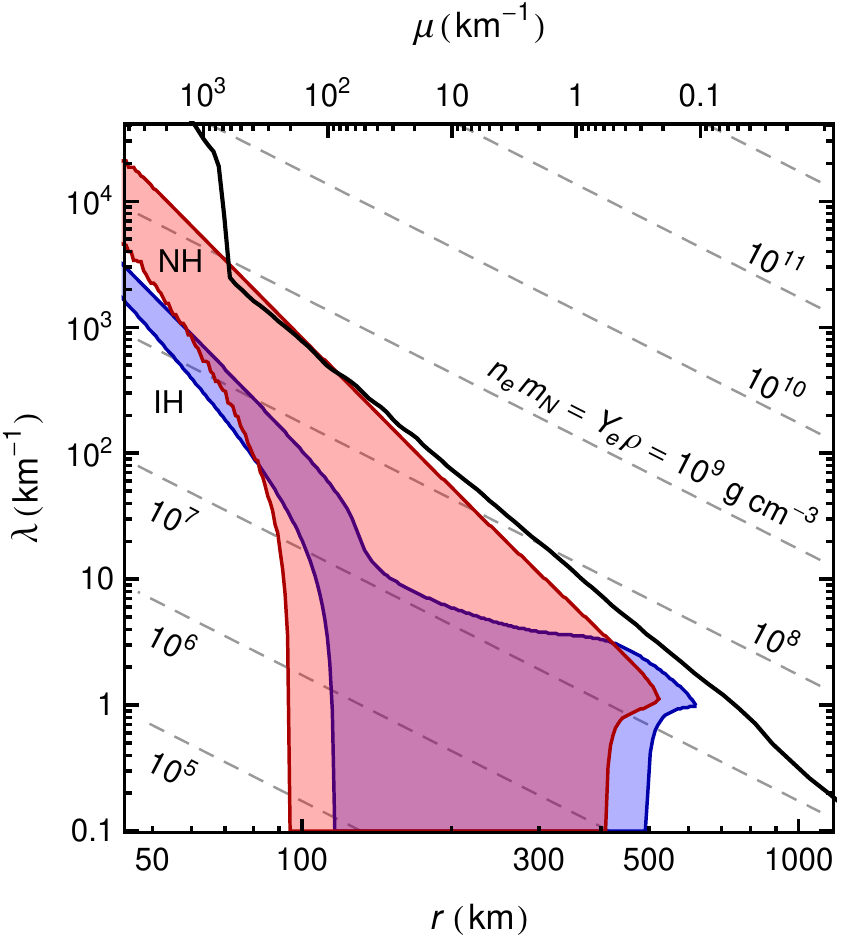}
\caption{Region where $\kappa r>1$ for IH (blue) and NH
  (red), depending on radius $r$ and multi-angle matter
  potential $\lambda$ for our simplified SN model.
  {\em Thick black line:} SN density profile.
  {\em Thin dashed lines:} Contours of constant electron density,
  where $Y_e$ is the electron abundance per baryon.
  (The IH case corresponds to Fig.~4 of Ref.~\cite{Sarikas:2011am},
  except for the simplified spectrum used here.)
  \label{fig:footprint}}
\end{figure}

{\em Schematic SN example.}---During the SN accretion phase, the
matter effect can be so large as to suppress collective flavor
conversions \cite{EstebanPretel:2008ni, Sarikas:2011am,
Chakraborty:2011gd}. In Fig.~\ref{fig:footprint} we juxtapose the
instability regions for the IH bimodal and the new NH MAA
instabilities for a simplified SN model.  We use single energy and
blackbody-like emission at the neutrino sphere, ignoring the halo
flux \cite{Sarikas:2012vb}. We choose physical parameters $R$,
$\mu(R)$, and $\epsilon$ that mimic the more realistic $15\,M_\odot$
accretion-phase model used in our previous study
\cite{Sarikas:2011am, Sarikas:2012vb}. We show the region where
$\kappa r>1$, i.e., where the growth rate is deemed ``dangerous.'' We
also show $\lambda(r)$, where the shock wave is seen at 70~km.

The matter profile never intersects the bimodal instability region,
i.e., this instability is suppressed everywhere in this specific SN
example. On the other hand, $\lambda(r)$ intersects the MAA
instability region just outside the shock wave. This simplified case
illustrates that the MAA instability can arise in SN models where
the bimodal instability is suppressed. It also shows that the
``danger spots'' are in very different places, although it remains
to be seen if this finding is generic.

{\em Conclusions.}---All previous studies of self-induced neutrino
conversion in SNe or the early universe were based on the false
premise that solutions of the equations of motion would inherit the
symmetries of the initial or boundary conditions. We have shown that
azimuth-angle instabilities are a generic phenomenon of collective
neutrino oscillations. Every single case in the previous literature
with enforced axial symmetry may have missed the dominant effect.

We have linearized the equations of motion around the initial state
of neutrinos in flavor eigenstates. The system then shows either the
bimodal or the MAA instability, but not both. (For more complicated
spectra that would lead to multiple spectral splits
\cite{Dasgupta:2009mg}, the bimodal instability occurs for positive
spectral crossings, the MAA instability for negative ones.) However,
evolved bimodal solutions, where the off diagonal $\varrho$ entries
are not small, may still become $\varphi$-unstable, and the other
way round.

Both instabilities can be suppressed by matter, but the required
density is larger for MAA. Therefore, it is not necessarily clear if
collective flavor conversions are generically suppressed during the
SN accretion phase, an important question for possible neutrino mass
hierarchy determination \cite{Serpico:2011ir}. For those cases where
suppression is not effective, dedicated numerical studies are
needed.

More fundamentally, one also needs to question the validity of other
common symmetry assumptions. For example, we have assumed a
stationary solution inherited from stationary neutrino emission.
Doubts may be motivated, in particular, by the role of the small
backward flux caused by residual neutrino scattering that causes
significant refraction \cite{Cherry:2012zw, Sarikas:2012vb}. Even
without worrying about the backward flux, it has never been proven
that a stationary boundary condition implies a stationary solution
for a dense interacting neutrino stream. In the early universe,
homogeneous initial conditions need not guarantee homogeneous
solutions. It remains to be seen if the interacting neutrino system
can spontaneously break translation symmetry in space or time.

{\em Note added.}--—Motivated by the preprint version of our paper, a
numerical study has appeared that confirms the existence and
importance of the MAA instability \cite{Mirizzi:2013rla}. Moreover,
two of us have devised a simple toy example of two
counter-propagating beams that shows a flavor instability in both
neutrino mass hierarchies and explains the physical nature of the MAA
instability \cite{Raffelt:2013isa}.


{\em Acknowledgments.}---We thank B.~Dasgupta, I.~Tamborra, and
H.-T.~Janka for comments on the manuscript. This work was partly
supported by the Deutsche Forschungsgemeinschaft under grant EXC-153
(Cluster of Excellence ``Origin and Structure of the universe'') and
by the European Union under grant PITN-GA-2011-289442 (FP7 Initial
Training Network ``Invisibles''). D.~S. acknowledges support by the
Funda\c{c}\~{a}o para a Ci\^{e}ncia e Tecnologia (Portugal) under
grant SFRH/BD/66264/2009.


\end{document}